\documentclass[journal,comsoc]{IEEEtran}
\usepackage[T1]{fontenc}

\usepackage{cite}

\ifCLASSINFOpdf
   \usepackage[pdftex]{graphicx}
   \graphicspath{{figure/}}
   \DeclareGraphicsExtensions{.pdf,.jpeg,.png}
\else
\fi
\usepackage{amsmath}
\usepackage{amssymb}

\usepackage[cmintegrals]{newtxmath}
\usepackage{bm}

\usepackage{array}

\ifCLASSOPTIONcompsoc
  \usepackage[caption=false,font=normalsize,labelfont=sf,textfont=sf]{subfig}
\else
  \usepackage[caption=false,font=footnotesize]{subfig}
\fi

\usepackage{stfloats}

\usepackage{multirow}

\begin{document}
%
\title{Deep Learning based Channel Estimation Algorithm over Time Selective Fading Channels}
%
%
%

\author{Qinbo~Bai,
        Jintao~Wang,~\IEEEmembership{Senior Member,~IEEE,}
        Yue~Zhang,~\IEEEmembership{Senior Member,~IEEE,}
        and~Jian~Song,~\IEEEmembership{Fellow,~IEEE}
\thanks{This work was supported in part by the National Key R\&D Program of China under Grant 2017YFE0112300 and Beijing National Research Center for Information Science and Technology under Grant BNR2019RC01014 and BNR2019TD01001.(Corresponding author: Jintao Wang.)}
\thanks{Qinbo Bai, Jintao Wang, and Jian Song are with the Electronic Engineering Department, Tsinghua University, and Beijing National Research Center for Information Science and Technology (BNRist), Beijing 100084, China (e-mail:wangjintao@tsinghua.edu.cn).}
\thanks{Yue Zhang is with the Department of Engineering, University of Leicester, Leicester, LE1 7RH, United Kingdom (e-mail: yue.zhang@leicester.ac.uk).}}

\maketitle

\begin{abstract}
The research about deep learning application for physical layer has been received much attention in recent years. In this paper, we propose a Deep Learning(DL) based channel estimator under time varying Rayleigh fading channel. We build up, train and test the channel estimator using Neural Network(NN). The proposed DL-based estimator can dynamically track the channel status without any prior knowledge about the channel model and statistic characteristics. The simulation results show the proposed NN estimator has better Mean Square Error(MSE) performance compared with the traditional algorithms and some other DL-based architectures. Furthermore, the proposed DL-based estimator also shows its robustness with the different pilot densities.
\end{abstract}


%
\IEEEpeerreviewmaketitle

\section{Introduction}
%
%
%
%
\IEEEPARstart{A}{s} the machine learning technology and the performance of hardware develop rapidly in recent years, Deep Learning(DL) has been successfully applied to many fields, especially in Computer Version and Nature Language Processing(NLP). Such technology has been applied to the physical layer processing of communication systems in \cite{8054694}. Since then, more research has been focusing on applying learning algorithms to different communication user scenarios.
\par
In traditional communication system, it always consists of different modules such as source coding, channel coding, modulation, demodulation, estimation, equalization, etc. 
And an end-to-end communication system under AWGN channel is designed in \cite{8054694}. Using fully connected NN, whose behavior is similar to an autoencoder, it achieves the similar performance to the tradition system with (7,4) Hamming code and BPSK modulation. Such autoencoder learns how to get an expression in a low dimension and the way to restore it. 
And Convolution Neural Network(CNN) based model \cite{8664650} has been developed to solve the dimensional explosion problem in autoencoder and achieves better performance than traditional methods(64QAM+MMSE) under both AWGN and static fading channel. 
Besides, a communication system with Software Defined Radio(SDR) only including NN are used to prove that transmission over the air with deep learning technology is possible \cite{8214233}. 
In Orthogonal Frequency Division Multiplexing \cite{387096} system, the deep learning algorithm for joint channel estimation and signal detection has been researched in \cite{8052521}. 
To overcome the back-propagation problem in NN transmitter when the channel is unknown, different methods are proposed. Policy gradient algorithm in the reinforcement learning is used in \cite{8433895}. 
A new deep learning technology, Conditional Generative Adversarial Nets\cite{CGAN}, is introduced in \cite{8644250} to emulate the unknown channel.
Simultaneous Perturbation Stochastic Approximation\cite{SPSA} algorithm is utilized in \cite{8452950} to give a direct estimation of the channel gradient.
\par However, in order to make DL-based communication system meaningful in the practical system, complex channels need to be considered. One kind of complex channel, which is difficult to handle with a traditional algorithm, is the time selective channel. Due to the movement of receiver, the channel status will change in time domain. Research on such channel using deep learning is somehow only a little. 
Sliding Bidirectional Recurrent Neural Network(SBRNN) has been put forward in \cite{8454325} and works as a detector to learn rapid varying optical and molecular channel. 
A simple application of neural network to Rayleigh fading channel is given in \cite{8554830}.  
Multiple Layers Perceptron(MLP) is used to undertake channel estimation for the time selective channel \cite{8491068} and doubly selective channel \cite{8672767}, respectively.
\par However, MLP is a memoryless structure. Thus, it can't learn the relation of data in time domain well. Besides, linear layers in MLP will result in the increasing of neurons size as input length increases. Despite that data can be divided into blocks to avoid this problem, divided data may lead to the discontinuity of the channel estimation. Considering the similarity of this problem in NLP field, it is better to use Recurrent Neural Network(RNN) to get the estimation of channel. In this article, time varying Rayleigh fading channel is explored using the deep learning technology and our contributions are summarized as below.
\begin{itemize}
	\item Based on deep learning algorithm, the SBGRU channel estimator is proposed to learn time varying Rayleigh fading channel. Using RNN structure and sliding idea, SBGRU can handle the transmitted symbol with arbitrary length and immediately provide the result as soon as the symbol arrived. 
	\item Substantial simulations are provided in this paper to analyze and explain NN estimator. The simulation result shows the ability of SBGRU to track channel dynamically and achieve better performance compared with traditional algorithms and other NNs. Besides, the SBGRU also has demonstrated the robustness with various pilot densities.
\end{itemize}
\par The rest parts of this article is arranged as follow: Section II describes the basic channel model, data structure and the signal flow model. Section III gives the deep learning based algorithm in details for NN channel estimator. Section IV uses quantities of simulation results to demonstrate the performance of NN estimator. Finally, Section V concludes the paper and gives some orientations for future work.
\par \textit{Notation}: Bold lower-case letters and upper-case letters denote vectors and matrices, respectively. The subscript on a lower-case letter $x_i$ represent $i^{th}$ element of vector $\bm{x}$. $E(\cdot)$ refers to the expectation. $(\cdot)^T$ and $(\cdot)^H$ refer to the transpose and Hermite transpose of the vector. $|\cdot|$ represents for the absolute value or amplitude for real number and complex number, respectively. For two vectors or matrices $\bm{a}$ and $\bm{b}$,$[\bm{a},\bm{b}]$ is the matrix combing $\bm{a}$ and $\bm{b}$. For two real numbers $a\leq b$, $[a,b]$ is the set for all real numbers in range from $a$ to $b$. $real(\cdot)$ and $image(\cdot)$ are the functions giving the real and imaginary part of complex vector for each element.

\section{System Model}
\par In this section, signal architecture and time varying Rayleigh fading channel model are firstly presented. Then, a signal flow model will be introduced. Denote the transmitted signal and received signal as $\bm{x}$,$\bm{y}$, respectively. Denote the Rayleigh time varying channel as $\bm{h}$. Considering a Linear Time Variant(LTV) model, the relation between input and output of channel is:
\begin{equation}
\label{E1}
\bm{y}=\bm{h}\cdot\bm{x}+\bm{\omega}
\end{equation}
where $\bm{\omega}$ is i,i,d Additive White Gaussian Noise(AWGN) vector, and $\omega_i\sim\mathcal{CN}(0,\sigma_n^2)$

\subsection{Time Varying Rayleigh Fading Channel Model}
\par Typically, wireless communication environment is generally modeled as Rayleigh fading channel. Multi-path will cause frequency selective fading and Doppler shifting will result in time selective fading. However, in this paper, only time selective fading is considered in order to give the first exploration of rapidly varying channel. The influence of multi-path will be researched in the future work.
\par Clarke's model \cite{6779222} is used in this paper to describe time varying channel. In order to describe the time varying characteristic, Jakes Doppler Spectrum\cite{1622098} is adopted here:
\begin{equation}
\label{E2}
S(f)=\frac{1}{\pi f_d\sqrt{1-(\frac{f}{f_d})^2}},\quad|f|<f_d
\end{equation}
where $f_d$ is the maximum Doppler shift. Given a speed $v$(m/s) and carrier frequency $f_c$(Hz), $f_d=\frac{vf_c}{c}$($c\approx3.0*10^8$ is the speed of light in free space). The autocorrelation of Jakes Doppler Spectrum is:
\begin{equation}
\label{E3}
R(\tau)=\int_{-f_d}^{f_d}S(f)\exp(j2\pi f\tau)df=J_0(2\pi f_d\tau)
\end{equation}
where $J_0(\cdot)$ is the first kind of Bessel function of 0 order and the discrete form of autocorrelation is:
\begin{equation}
\label{E4}
R[d]=J_0(2\pi\phi_d|d|)
\end{equation}
where $\phi_d=\frac{f_d}{r_s}$ is the maximum Doppler frequency normalized by sampling rate.
Besides, It is generally asked that the channel has normalized gain $E(|h[n]|^2)=1$ in order to simplify following analysis. 

\subsection{Signal Architecture}
\par
\begin{figure}[!t]
\centering
\includegraphics[width=3.5in]{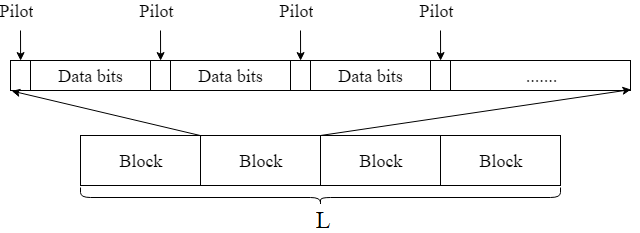}
\caption{Data structure of transmitted signal in time domain}
\label{img1}
\end{figure}

 Considering standard signal architecture, transmitted signals are generated as shown in Fig. \ref{img1}. One single data frame consists of $K$ blocks. Due to multi-path not considering in this article, protection interval isn't necessary. Each block has $N_s$ information symbols and $N_p$ pilot symbols. Thus, each block has $N_s+N_p=N$ symbols and the whole frame has total $L=NK$ symbols. Pilots are equally interval inserted in each block, and define $\frac{N_p}{N_s+N_p}$ as the pilot density. Besides, pilots in each block are the same, which results in repetition in time domain.

\subsection{Signal Flow Model}
\begin{figure}[!t]
	\centering
	\includegraphics[width=3.5in]{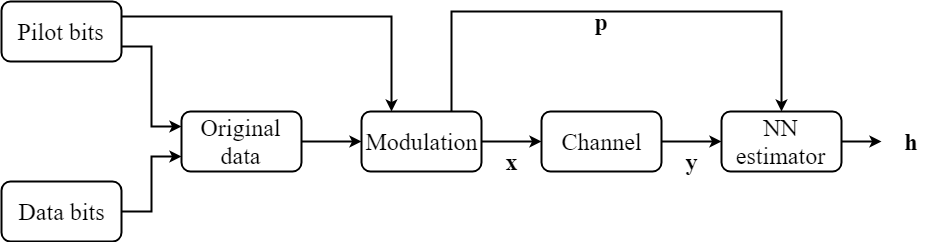}
	\caption{Signal flow model}
	\label{img2}
\end{figure}

The signal flow model is shown in Fig. \ref{img2}. At the transmitter side, no deep learning technology is introduced. Information bits and pilot bits are combined to generated original signal. After modulating, transmitted signal $\bm{x}$ is sent to the channel and modulated pilots $\bm{p}$ are sent to NN estimator. At the receiver side, NN channel estimator uses $\bm{p}$ and channel distorted signal plus the noise $\bm{y}$ to give the estimation of channel $\bm{h}$. 
\par Two things need to be notified. Firstly, due to no NN introduced at transmitter, it is easy to add any traditional channel coding such as Low Density Parity Check(LDPC)\cite{LDPC}, to improve the performance against noise. Secondly, NN channel estimator doesn't need any information about the channel. It means that the communication system is model free.

\subsection{Traditional Algorithms For Channel Estimation}
In channel estimation, the most common estimators are Least Square(LS)\cite{LS} estimator and Minimal Mean Square Error(MMSE)\cite{MMSE} estimator. According to (\ref{E1}), LS estimator under the time varying channel is:
\begin{equation}
\label{E5}
\hat{\bm{h}}_{LS}=\frac{\bm{y}}{\bm{x}}
\end{equation}
For those positions where pilots are inserted, above equation can be directly used to get the estimation. For other positions, linear interpolation is necessary. Denote $j$,$k(j<k)$ to be positions of pilot nearest to the position $i$. Thus, the interpolated channel is:
\begin{equation}
\label{E6}
\hat{h}_{i,LS}=\frac{k-i}{k-j}\hat{h}_{j,LS} + \frac{i-j}{k-j}\hat{h}_{k,LS}
\end{equation}
Due to the existence of noise, omitting the influence of interpolation, the expected Mean Square Error(MSE) of LS estimator is:
\begin{equation}
\label{E7}
E(|\hat{\bm{h}}_{LS}-\bm{h}|^2)=E(\frac{\bm{\omega}^2}{\bm{x}^2})=\frac{1}{SNR}
\end{equation}
\par Another traditional estimator would be MMSE estimator:
\begin{equation}
\label{E8}
\hat{\bm{h}}_{MMSE}=\bm{R}_{\bm{hy}}\bm{R}_{\bm{yy}}^{-1}\bm{y}=\bm{R}_{\bm{hh}}(\bm{R}_{\bm{hh}}+\frac{\sigma_n^2}{\sigma_s^2}\bm{I})^{-1}\hat{\bm{h}}_{LS}
\end{equation}
where $\bm{I}$ represents unit matrix and $\bm{R}_{\bm{hh}}=E(\bm{h}\bm{h^H})$ represents correlation matrix:
\begin{equation*}
\bm{R}_{\bm{hh}}=
\left[\begin{array}{ccccc}
R[0] & R[1] & R[2] & \cdots & R[L-1]\\
R[1] & R[0] & R[1] & \cdots & R[L-2]\\
R[2] & R[1] & R[0] & \cdots & R[L-3]\\
\vdots & \vdots & \vdots & \ddots & \vdots\\
R[L-1] & R[L-2] & R[L-3] & \cdots & R[0]\\
\end{array} 
\right]
\end{equation*}
where $R[\cdot]$ can be calculated according to (\ref{E4})
\par It should be noticed that the form of autocorrelation function of channel and Doppler speed need to be given in advance in order to undertake the MMSE estimation. However, real channel model and accurate statistic characteristic(Doppler speed here) are hard to know under practical application. Thus, two methods for MMSE estimation are used in simulation. 
\par Firstly, assuming above information already known, $\hat{\bm{h}}_{MMSE}$ can be directly calculated according to (\ref{E3}) and (\ref{E8}). Thus, we call this method "MMSE theory". Secondly, after getting LS estimation, $\hat{\bm{h}}_{LS}$ can be used to calculate autocorrelation $R[d]=\sum_{n=0}^{L-1}\hat{h}_{LS}[n]\hat{h}_{LS}[n-d]$ and then use (\ref{E8}). We call this method "MMSE sim" because the computation is completed by simulation results.

\section{DL-based NN channel estimator}
\par 
To track a time varying channel, it is necessary to give neural network the ability of studying the behavior of correlation in time domain. Thus, a good choice to handle sequence data is using RNN. 

\subsection{RNN structure}
\begin{figure*}[!t]
	\centering
	\subfloat[]{\includegraphics[width=3in]{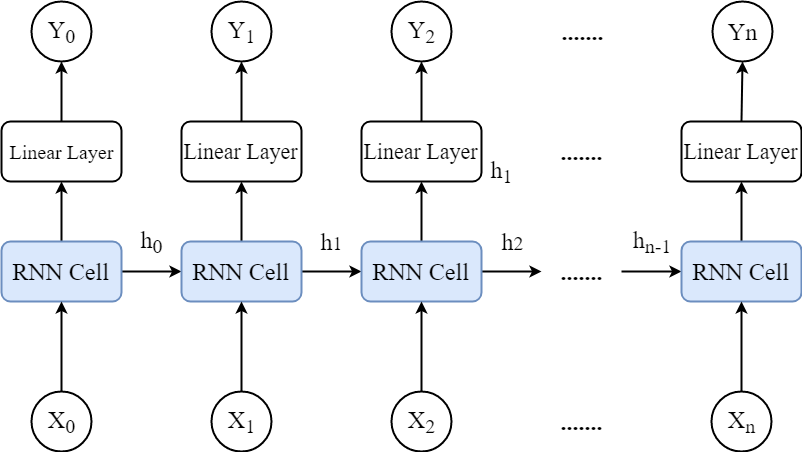}%
		\label{img3a}}
	\hfil
	\subfloat[]{\includegraphics[width=3in]{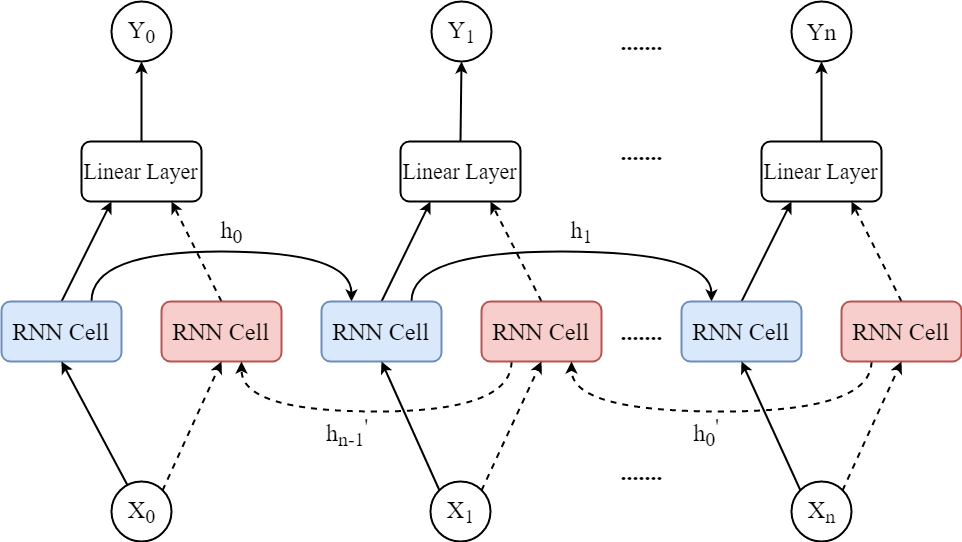}%
		\label{img3b}}
	\caption{The structure of RNN (a) The structure of forward only RNN (b) The structure of bidirectional RNN}
	\label{img3}
\end{figure*}
A simple example of 1 layer RNN is given in Fig. \ref{img3a}. In this structure, the output of last time becomes one part of input of this time. By this way, RNN can capture past information. The basic RNN cell will give the computation result as the following function.
\begin{equation}
\label{E9}
\bm{h}_t=Tanh(\bm{W}_{ih}\bm{x}_t+\bm{b}_{ih}+\bm{W}_{hh}\bm{h}_{t-1}+\bm{b}_{hh})
\end{equation}
where $Tanh$ is hyperbolic tangent function and $\bm{h}_t,\bm{h}_{t-1}$ are the hidden states at time $t$ and $t-1$, respectively. $\bm{x}_t$ is the input at time $t$. $\bm{W}_{ih},\bm{W}_{hh}$ and $\bm{b}_{ih},\bm{b}_{hh}$ are weights and biases, which need to be learned.
\par However, the time varying channel $h(t)$ has relation with both past and future channel states. Basic RNN cell is fed forward only. Thus, bidirectional structure, as shown in Fig. \ref{img3b}, would have better performance. Blue blocks are forward cells and red blocks are backward cells. The data will not only be fed in forward direction, but fed backward again. The hidden states $\bm{h}_t$ and $\bm{h}^{'}_t$ are combined together to become the input of a linear layer to give final results.
\par Another problem is that Basic RNN cell with (\ref{E9}) can't capture long time information. To solve this problem, Long Short Time Memory(LSTM)\cite{LSTM} cell has been put forward. In this paper, Gated Recurrent Unit(GRU)\cite{GRU} is used, one variation of LSTM, to replace basic RNN cell. The GRU will give the result as the following function(\cite{GRU},(5),(6),(7),(8))
\begin{subequations}
	\label{E10}
	\begin{equation}
	\label{E10a}
	\bm{z}_t=\sigma(\bm{W}_z\cdot[\bm{h}_{t-1},\bm{x}_t])
	\end{equation}
	\begin{equation}
	\label{E10b}
	\bm{r}_t=\sigma(\bm{W}_r\cdot[\bm{h}_{t-1},\bm{x}_t])
	\end{equation}
	\begin{equation}
	\label{E10c}
	\overline{\bm{h}}_t=Tanh(\bm{W}\cdot[\bm{r}_t*\bm{h}_{t-1},x_t])
	\end{equation}
	\begin{equation}
	\label{E10d}
	\bm{h}_t=(1-\bm{z}_t)*\bm{h}_{t-1}+\bm{z}_t*\overline{\bm{h}}_t
	\end{equation}
\end{subequations}
where $\sigma(\cdot)$ refers to Sigmoid function $f_s(x)=\frac{1}{1+e^{-x}}$, $\bm{W}_z,\bm{W}_r,\bm{W}$ are weights and $\bm{h}_t,\bm{h}_{t-1},\bm{x}_t$ have the same meaning as (\ref{E9}).
Compared with basic RNN cell, GRU introduces 2 gates, update gate $\bm{z}_t$ and reset gate $\bm{r}_t$, to control the information flow. GRU has been proved to have similar performance to LSTM on many tasks\cite{7508408} and have higher speed due to less gate number.
\par Based on above discussion, BGRU cell will be used in NN channel estimator. However, the result of simple BGRU is not good enough. The idea of Sliding BRNN(SBRNN)\cite{8454325} is considered to improve the performance further, and the compare between BGRU and SBGRU will be given in section IV.C.

\subsection{SBGRU structure}
\par
\begin{figure}[!t]
	\centering
	\includegraphics[width=3.5in]{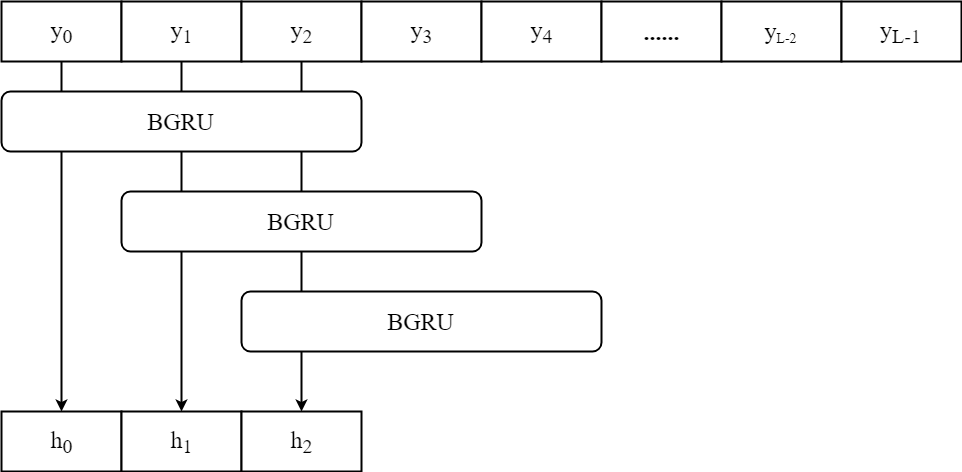}
	\caption{Sliding structure of BGRU}
	\label{img4}
\end{figure}

 SBRNN is put forward in \cite{8454325} to work as a detector under optical and molecule channel. Here, this structure is used in estimation task under the time varying Rayleigh fading channel. A simple example of the sliding structure is given in Fig. \ref{img4}. 
Each BGRU block in the figure has a fixed window length $W_L$. It should be stated that the selection of window length has relationship with channel character. Due to the any two moments of channel $h$ is correlated, it is reasonable that the longer the window is, the better the performance will be. The simulation about window length will be given in section IV.D.
\par SBGRU will be given $W_L$ symbols to undertake once computation, and will slide 1 symbol after each computation. Due to the sliding operation, most symbols in the sequence will be estimated for several times. We take the average of all estimation to give final results. Denote $\bm{h}_t=f_{BGRU}(\bm{x_t},\bm{h}_{t-1},\bm{h'}_{t+1})$ as the function of operation defined in (\ref{E10}) for bidirectional version in BGRU layer. Denote $\mathbf{S}=\{j|j\in\mathbb{Z},max(0,t-W_L+1)\leq j\leq min(t,L-1)\}$ as the set including all starting positions of BGRU for symbol $\bm{x}_t$, and final output of SBGRU for $\bm{x}_t$ is:
\begin{equation}
\label{E11}
\bm{h}_t=\frac{1}{|\mathbf{S}|}\sum_{j\in\mathbf{S}}f_{BGRU}(\bm{x_t},\bm{h}_{t-1}^j,\bm{h}_{t+1}^{'j})
\end{equation}
where $\bm{h}_{t-1}^j$ and $\bm{h}_{t+1}^{'j}$ are the hidden states for BGRU starting from $j^{th}$ symbol in time $t-1$ for forward and $t+1$ for backward.

\subsection{Train and test NN estimator}
\begin{figure}[!t]
	\centering
	\includegraphics[width=3.2in]{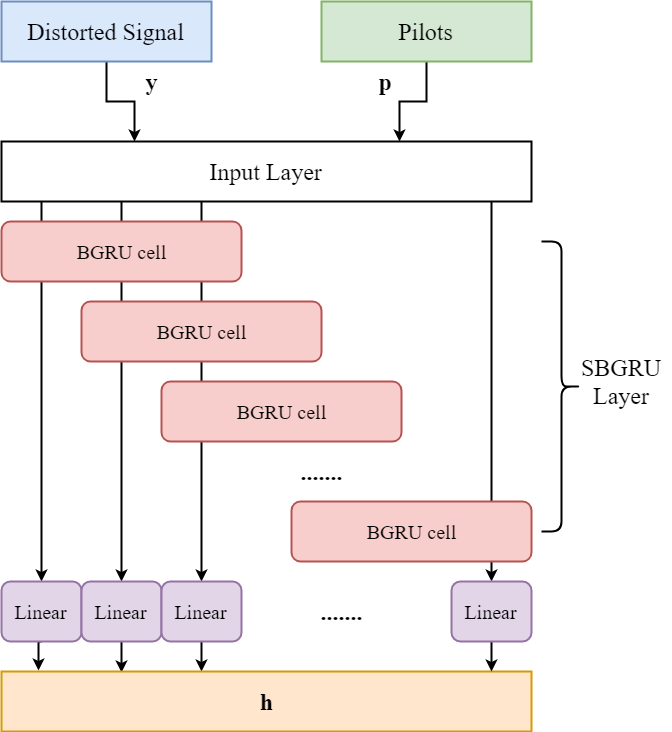}
	\caption{The structure of SBGRU channel estimator}
	\label{img5}
\end{figure}
\par A final implement of SBGRU neural network is given in Fig. \ref{img5}. The input data to SBGRU consists of channel distorted signal $\bm{y}$ and the original pilot information $\bm{p}=[\bm{p}^1,\bm{p}^2,...,\bm{p}^K]$, including $K$ same pilot blocks $\bm{p}^i$ with length $N$, and
$$\bm{p}^i=[p_1,0_{1*N_s},p_2,0_{1*N_s},...,p_{N_p},0_{1*N_s}]$$
It means that the pilot sequence will have the same symbols as $\bm{x}$ in pilot positions and have 0 symbols in information positions. Due to current deep learning platform only receiving real numbers, real part and image part of complex signal need to be separated firstly. Thus, the input data of the SBGRU will be given as:
$$\bm{X}_{in}=[real(\bm{y}^T),real(\bm{P}^T),imag(\bm{y}^T),imag(\bm{P}^T)]^T$$
\par  Considering the balance between  accuracy and training time, here 2 layers BGRU are adopted to construct SBGRU layer.
Denote function $f_{SBGRU}$ as the operation in SBGRU layer defined by (\ref{E11}) and function $f_{Linear}$ as the operation in Linear layer defined as:
\begin{equation}
\label{E12}
 f_{Linear}(\bm{x})=\bm{W}\bm{x}+\bm{b}
\end{equation}
where $\bm{W},\bm{b}$ are weight and bias in linear layer, respectively.
The final estimation of channel, denoted as $\hat{\bm{h}}$, can be expressed as:
\begin{equation}
\label{E13}
\hat{\bm{h}}=f_{Linear}(f_{SBGRU}(\bm{X}_{in},\bm{\theta}_{S}),\bm{\theta}_{L})
\end{equation}
where $\bm{\theta}_{S}$ are the parameters of SBGRU and $\bm{\theta}_{L}$ are the parameters of Linear layer.
\par Denote $\bm{\theta}=\{\bm{\theta}_{S},\bm{\theta}_{L}\}$ to make notation clearly. To train the NN estimator, a loss function, which can represent the system performance, needs to be constructed. And parameters $\bm{\theta}$ need to be optimized in order to minimize the loss function. Due to MSE always regarded as criterion in estimation problem, MSE loss function is adopted, which can be expressed as
\begin{equation}
\label{E14}
Loss(\theta)=\frac{1}{L}\sum_{n=1}^{L}|\hat{h}_n-h_n|^2
\end{equation}
Minimizing loss function can be completed by updating $\bm{\theta}$ iteratively. The most classical algorithm is Stochastic Gradient Descant(SGD). Adam\cite{Adam} optimization algorithm, which has better performance in multiple tasks, is adopted here.
\par Testing data has the same structure and statistic characteristics with training data. Trained parameters $\bm{\theta}$ are loaded to finish the computation of testing data and get the estimated channel.

\section{Simulation Results}
In this section, we demonstrate the performance of NN channel estimator under the time varying Rayleigh fading channel and provide the explanation to the performance improvement through the simulation results. And the simulation setting for the NN estimator is firstly described. Then, four group simulation results of NN estimator have been presented and analyzed.
\subsection{Simulation Setting}
In the following simulations, i.i.d. bit sequences are randomly generated, and QPSK modulation is used to map bits to symbols. According to the channel model given in section II.A and channel parameters given in Table \ref{Tab1}, 1200 channels are generated, 800 for training, 200 for validation and 200 for testing. The selection of channel parameters and pilot density is the same as \cite{8491068} in order to undertake  comparison simulation in Section IV.C. Also, based on the data structure in Fig. \ref{img1} and data parameters in Table \ref{Tab1}, 120000 sequences are generated, 100000 for training, 10000 for validation and 10000 for testing. When calculating the channel distorted signal, each symbol sequence randomly choose one channel to send.
\begin{table}[!t]
\renewcommand{\arraystretch}{1.3}
\caption{Channel and data parameters}
\label{Tab1}
\centering
\setlength{\tabcolsep}{12mm}{
\begin{tabular}{c|c}
\hline
Carrier frequency & 5.2GHz\\
\hline
Sampling rate & 0.25MHz\\
\hline
Receiver speed & 10m/s\\
\hline
Signal Length & 160 symbols\\
\hline
Pilot density & 50\% \\
\hline
Signal Block Length & 16 symbols\\
\hline
Signal Block Number & 10 symbols\\
\hline
\end{tabular}}
\end{table}
\par The default data and NN parameters of estimator, detector and system are shown in Table \ref{Tab2}.
\begin{table*}[!t]
	\renewcommand{\arraystretch}{1.3}
	\caption{NN Parameters for simulation}
	\label{Tab2}
	\centering
	\setlength{\tabcolsep}{20mm}{
	\begin{tabular}{c|c}
		\hline
		Parameter & Estimator \\
		\hline
		NN architecture & SBGRU \\
		\hline
		Number of hidden layers & 2\\
		\hline
		Hidden size & 40*2(2 for Bi-direction) \\
		\hline
		Window Length & 40 symbols\\
		\hline
		Activation function & Tanh for hidden layers \& Relu for hidden layers \\
		\hline
		Loss function & MSE \\
		\hline
		Optimizer & Adam \\
		\hline
		Learning rate & 0.001 \\
		\hline
		Batch Size & 128 \\
		\hline
		Train SNR & 20dB \\
		\hline
		Test SNR & 5,10,15,20,25dB \\
		\hline
		Train number & 100000 \\
		\hline
		Validation number & 10000\\
		\hline
		Test number & 10000 \\
		\hline
	\end{tabular}}
\end{table*} 
\par The proposed DL-based algorithm is implemented on a computer with an Intel (R) Corel (TM) i7-6700K CPU @ 4.0GHz CPU, a NVIDIA GeForce GTX 1080 GPU and 16GB memory. Pytorch 1.0.0 and python 3.6 are used for the estimation.

\subsection{Performance Comparison with the traditional algorithm}
\par
\begin{figure}[!t]
	\centering
	\includegraphics[width=3.5in]{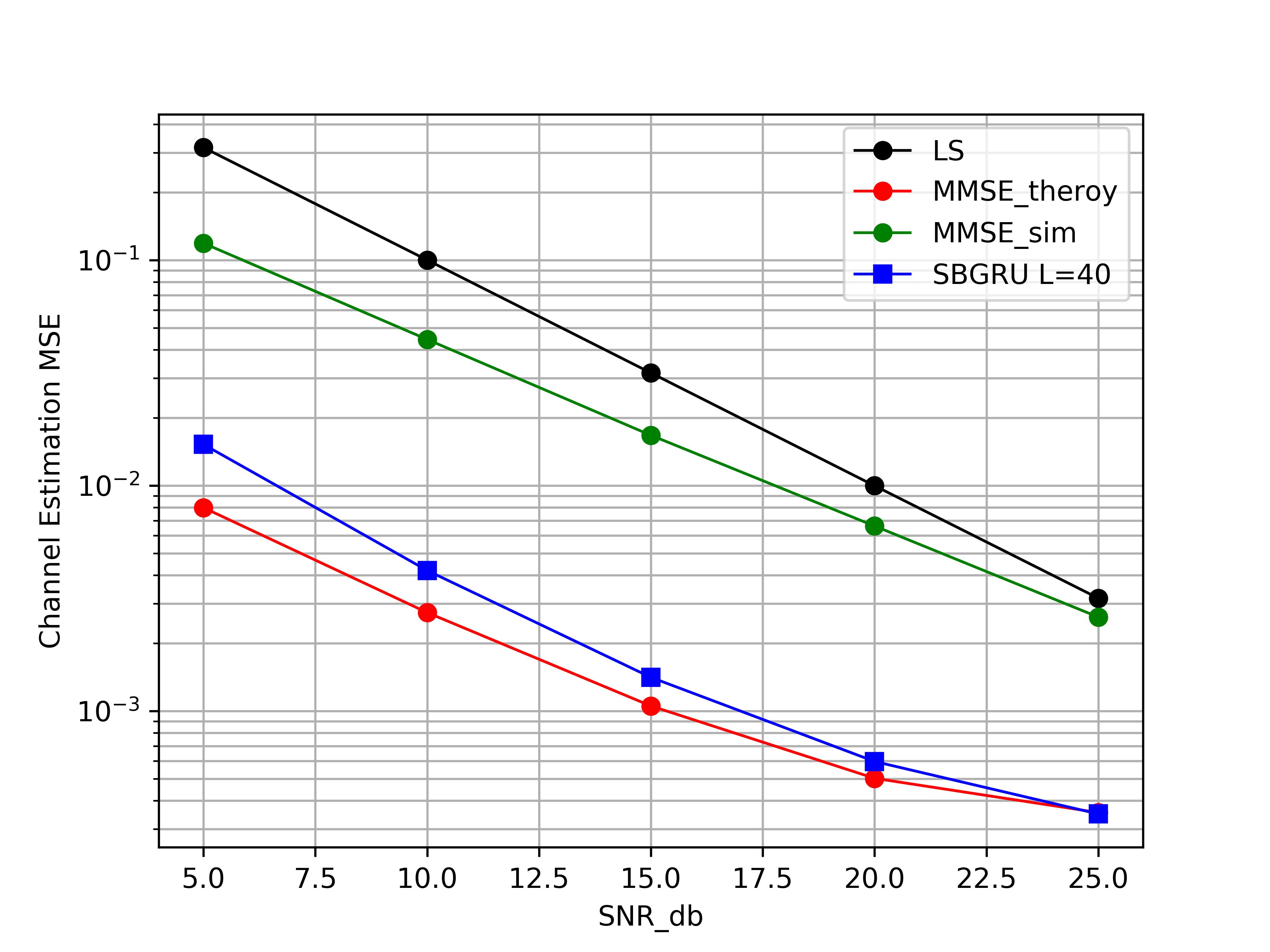}
	\caption{Performance compare with LS and MMSE estimator}
	\label{img6}
\end{figure}

 Here the proposed NN channel estimator is compared with traditional algorithm, LS estimator and MMSE estimator. The performance comparison is shown in Fig. \ref{img6}. It is obvious that "MMSE theory" achieves the best performance within the testing SNR range. And the LS estimation is the worst due to not considering the influence of the noise. And the simulation result does match the expected performance stated in (\ref{E7}). "MMSE sim" estimation, stated in section II.E, has small performance improvement compared with LS estimator and the improvement decreases when SNR reaches high value. SBGRU estimator reaches the similar performance to "MMSE theory" estimator and doesn't need any channel knowledge. Besides, SBGRU estimator also greatly outperforms both LS and "MMSE sim". Such results prove that SBGRU estimator is a best solution under the time varying channel.
\par
\begin{figure*}[!t]
	\centering
	\subfloat[]{\includegraphics[width=3.5in]{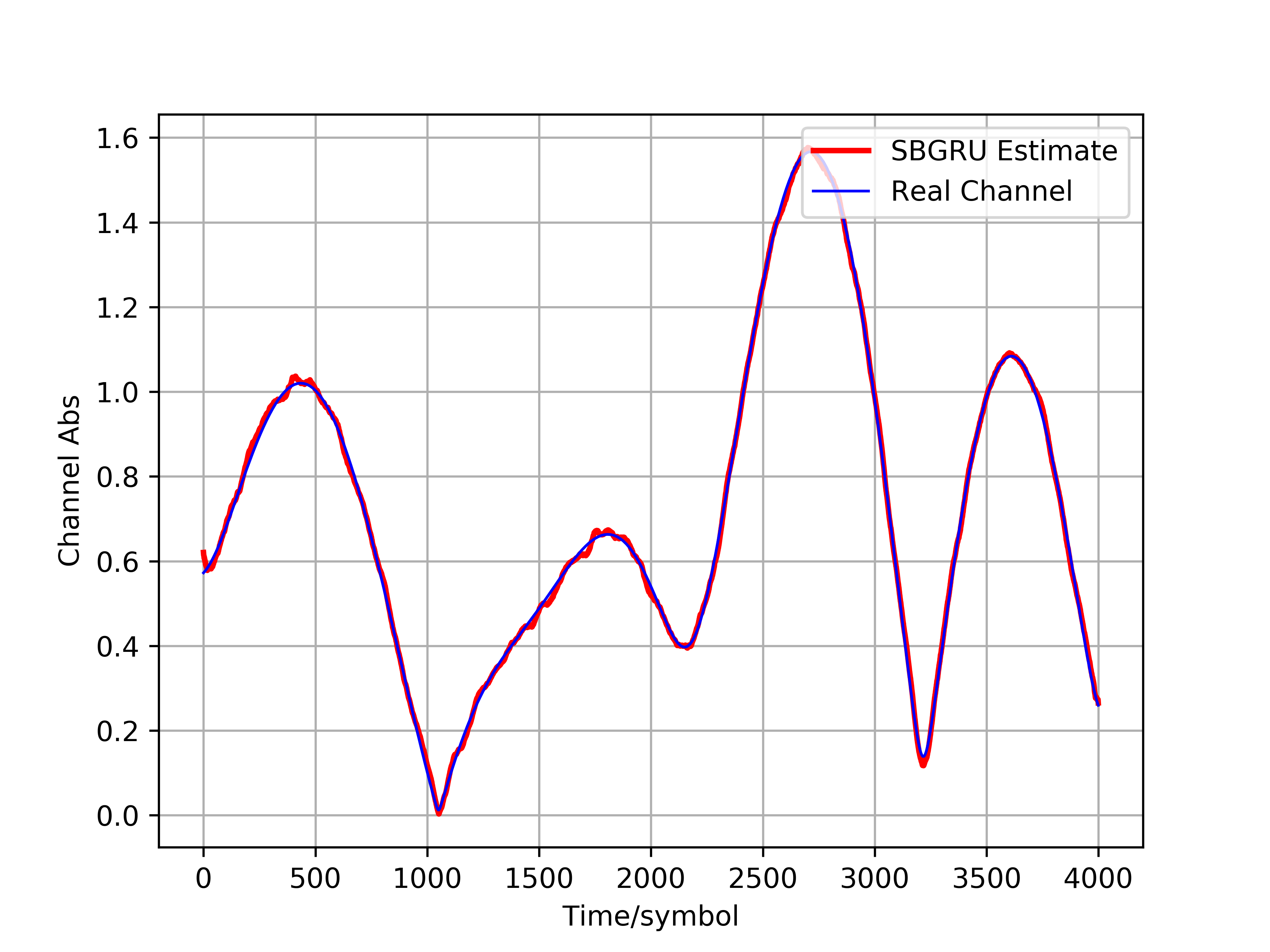}%
		\label{img7a}}
	\hfil
	\subfloat[]{\includegraphics[width=3.5in]{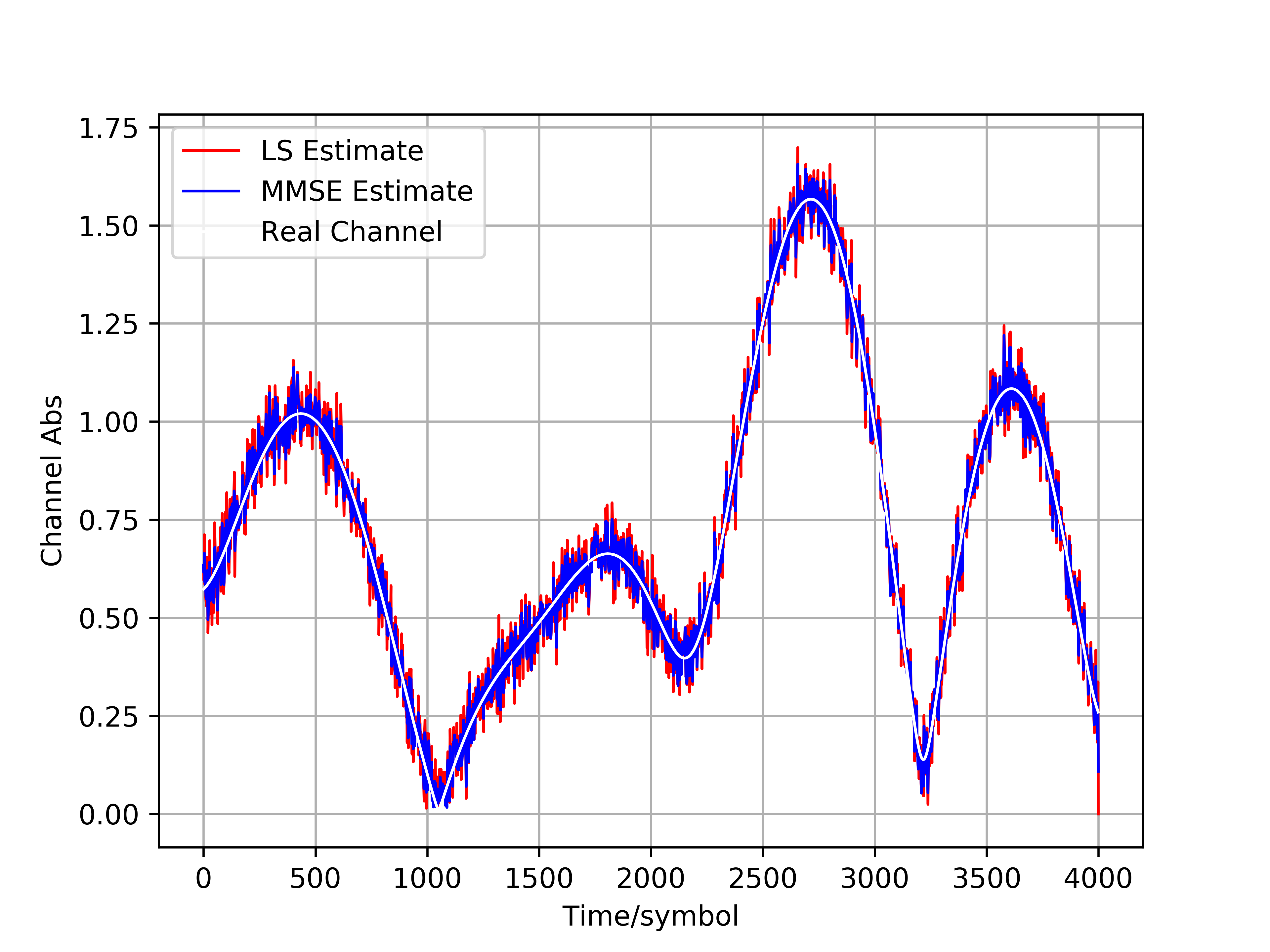}%
		\label{img7b}}
	\caption{Simulation results for Channel Tracking. (a) Tracking performance of SBGRU estimator (b) Tracking performance of LS and MMSE estimator}
	\label{img7}
\end{figure*}

 To visualize how the SBGRU estimator work, the performance of the channel tracking of the SBGRU and traditional estimator is given in Fig. \ref{img7}. In order to make the channel varying significant in time domain, channel length is extended to 4000 symbols and SNR is set to 20dB. It's easy to find that SBGRU estimator can track the channel very well in most linear parts and has slight oscillation in non-linear parts. However, in Fig. \ref{img7b}, where white line represents real channel, both LS estimator and "MMSE sim" estimator vibrate heavily. 

\subsection{Performance Comparison with different structures of NN}
\par
\begin{figure}[!t]
	\centering
	\includegraphics[width=3.5in]{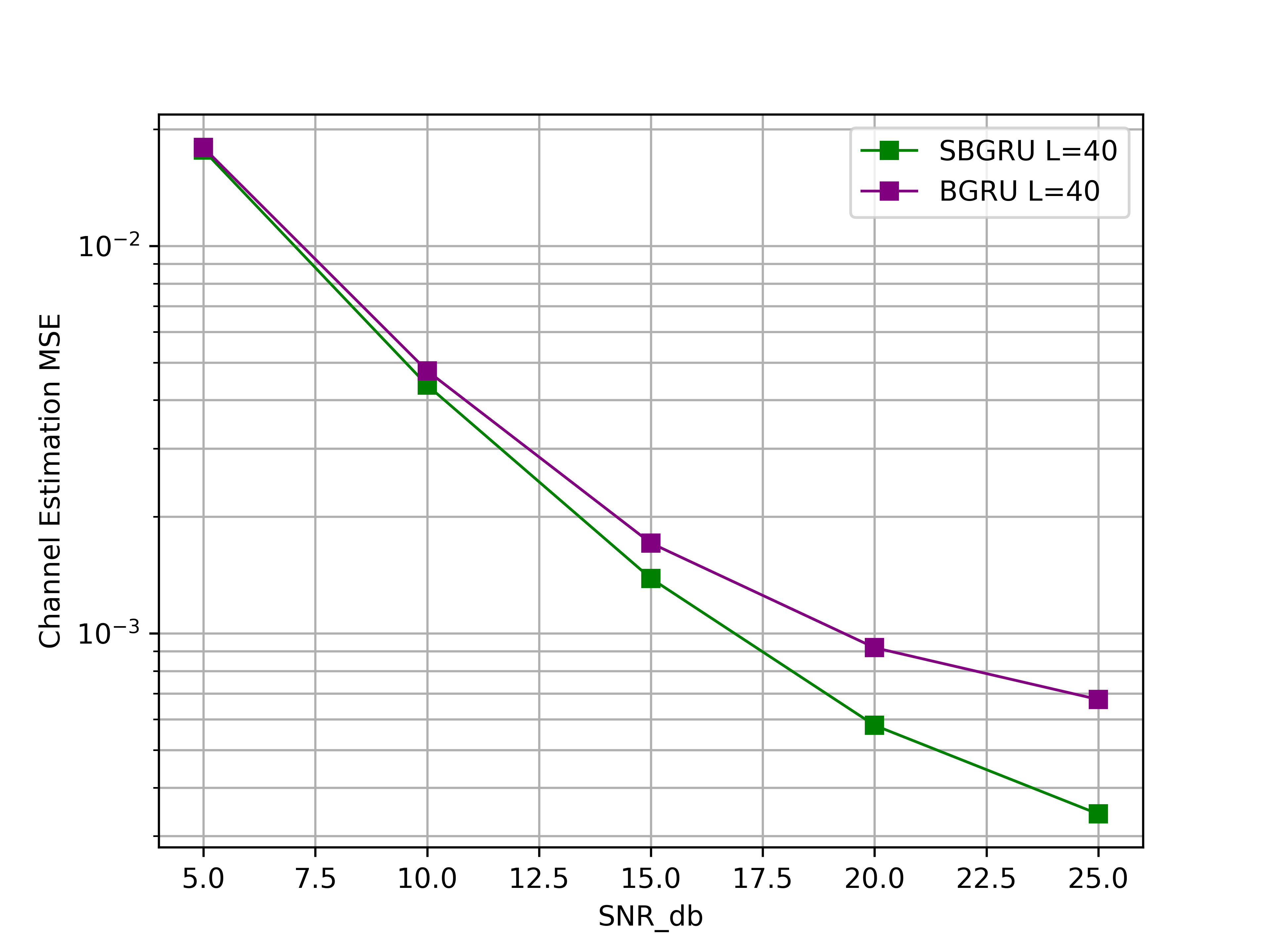}
	\caption{Performance compare between Sliding BGRU and Non-sliding BGRU}
	\label{img8}
\end{figure}

 When deep learning algorithms are used to undertake the channel estimation, different structures of neural network will achieve different performances. Firstly, The enhancement of the sliding operation for SBGRU is demonstrated in Fig. \ref{img8}. All settings are the same except that BGRU computes block by block. The performance of BGRU decreases rapidly as SNR increases because the introduction of sliding operation can utilizes the average channel information within a certain time window.
\par Besides, the channel estimation problem under similar time varying channel has been researched in \cite{8491068} by using MLP neural network. Its basic idea is to include not only the channel distorted data and pilot data but the estimated channel from last block to get the better channel estimation performance. In its simulations, it sets the estimation block length the same as the data structure. However, this estimation block length can be different. In order to compare the performance fairly, the NN architecture in \cite{8491068} is reconstructed, trained and tested using the same settings and simulation parameters as the SBGRU simulation. Besides, three different parameters $16$, $32$ and $40$ are used to fully explore the influence of the estimation block length, .
\par
\begin{figure}[!t]
	\centering
	\includegraphics[width=3.5in]{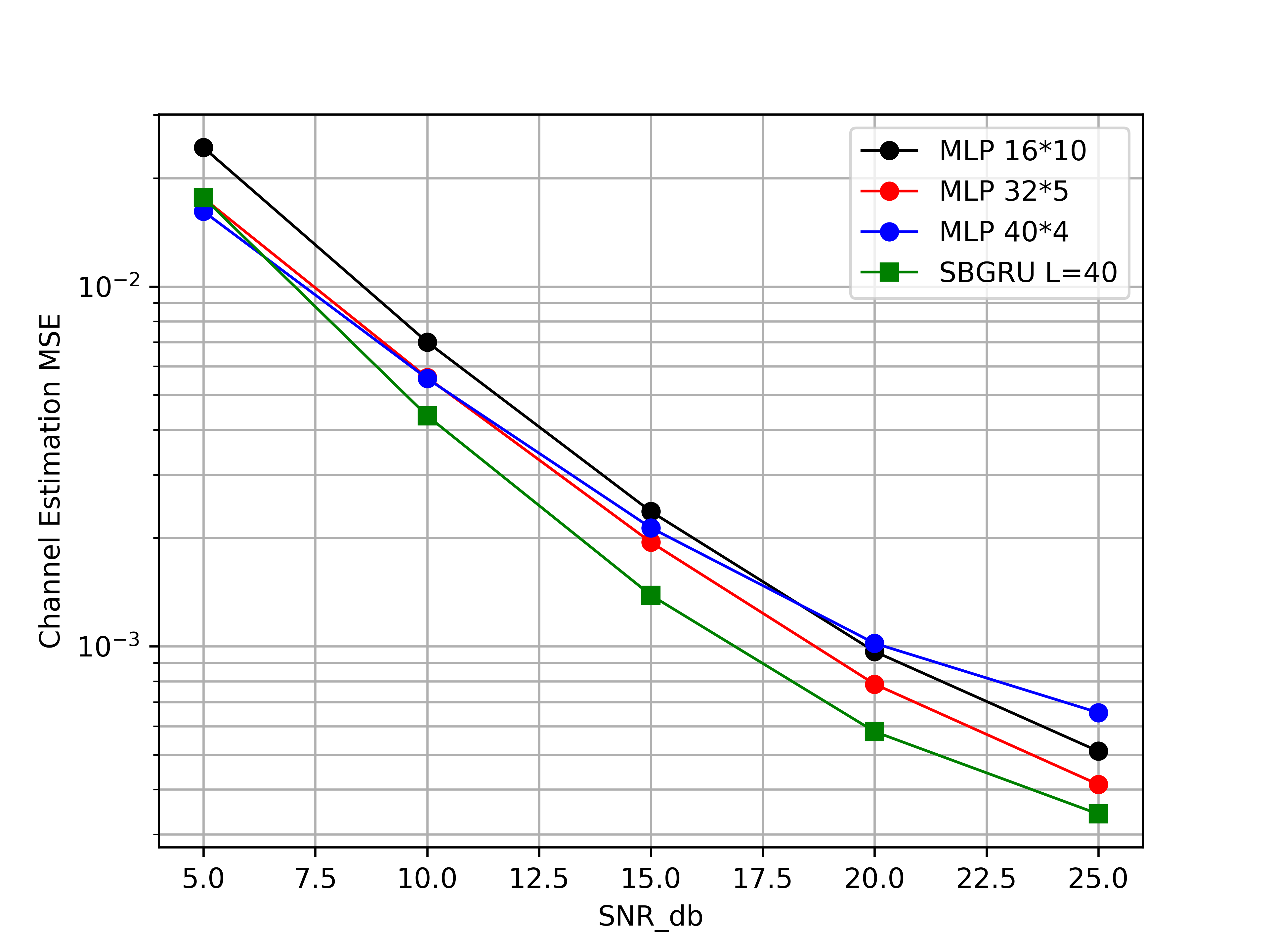}
	\caption{Performance compare between SBGRU estimator and MLP estimator}
	\label{img9}
\end{figure}

 The performance comparison between MLP and SBGRU is given in Fig. \ref{img9}. MLP with estimation block length 16(same design as \cite{8491068}) doesn't work very well. It is possible that parameters in NN model is not enough so that the ability to learn the nonlinear channel isn't strong. When estimation block length increases to 32, the performance increase a bit. However, a estimation block length of 40 will result in performance decreased. It is because MLP with estimation block length 40, which is not the integral multiple of original data block length 16, can't fully explore the pilot information repeated in time domain. However, SBGRU estimator outperforms all above MLP estimator when SNR is above 5dB. Besides, thanks to the recurrent structure of RNN, previous channel estimation doesn't need to be inputed into neural network. It can be captured by SBGRU automatically.

\subsection{Performance vs window length}
\par
\begin{figure}[!t]
	\centering
	\includegraphics[width=3.5in]{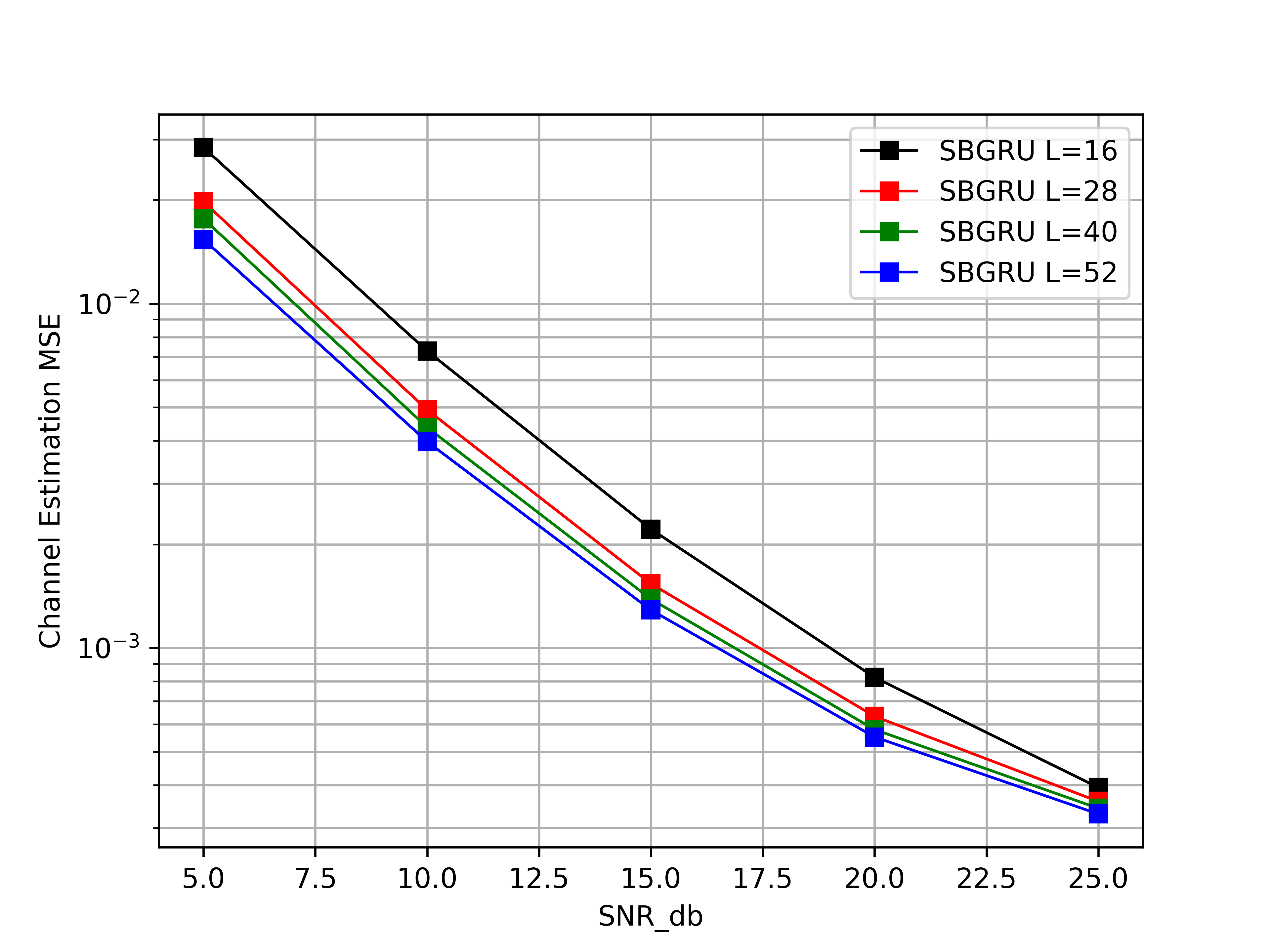}
	\caption{The influence of sliding window length to SBGRU estimator}
	\label{img10}
\end{figure}

 Here the influence of sliding window length is explored. The performance among different window lengths is in Fig. \ref{img10}. The performance monotonically increases as the window length getting longer. Except for window length of 16 symbols, all 3 other window lengths have nearly the same performance. It shows that the window length can't be too short in order to have enough information to undertake the estimation. However, the too long window length can't bring much more improvement. Thus, selecting a suitable window length can achieve the balance between the accuracy and the speed of training and testing. Overall, the setting of window length have the relation with channel characteristics.

\subsection{Performance vs pilot density}
\par
\begin{figure}[!t]
	\centering
	\includegraphics[width=3.5in]{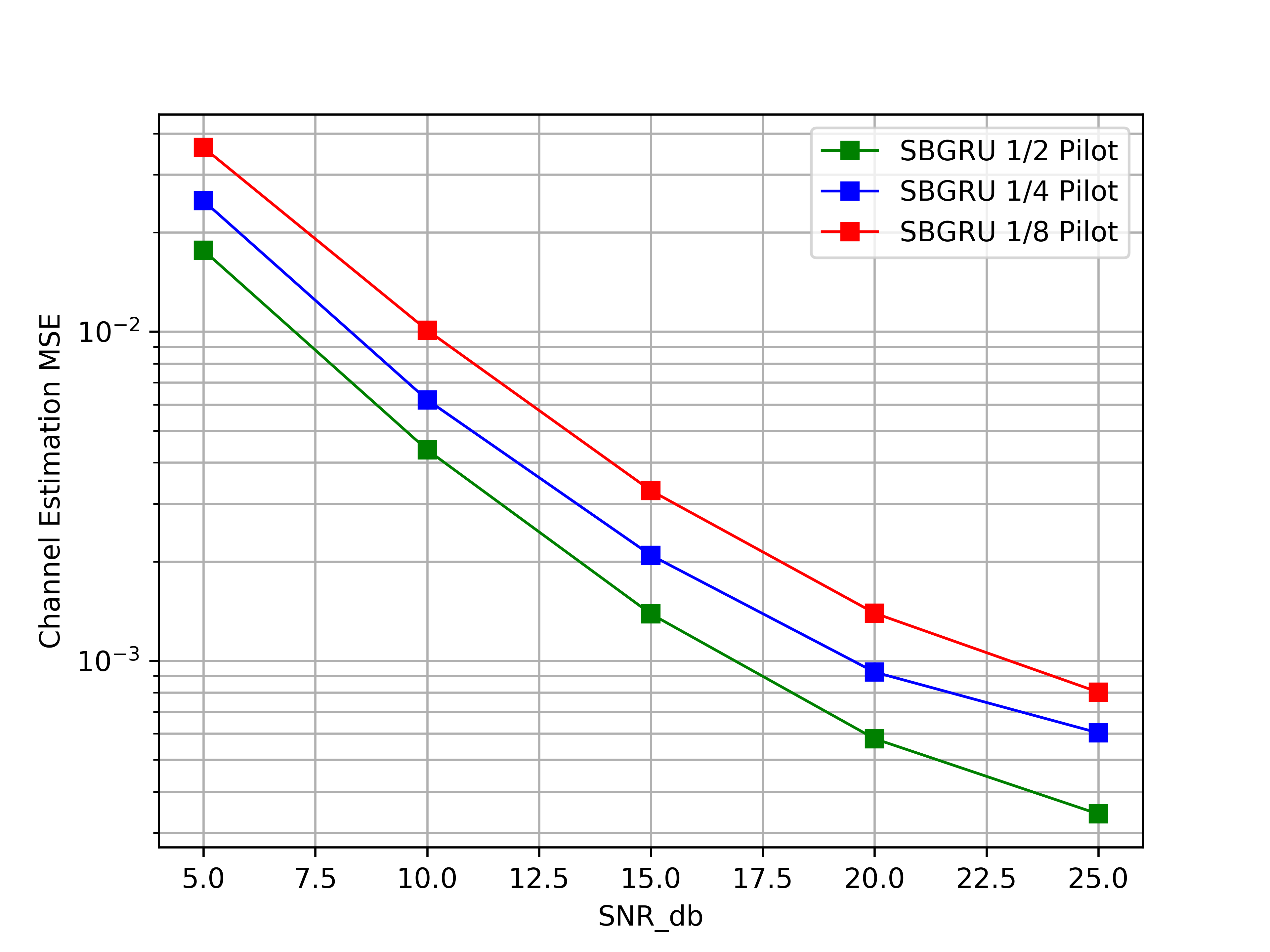}
	\caption{The influence of pilot density to SBGRU estimator}
	\label{img11}
\end{figure}

 Finally, the influence of pilot density is described to show the robustness of SBGRU estimator. The performance is shown in Fig. \ref{img11}. As the pilot density decreases, the MSE performance indeed decreases a little but not seriously. The result is still much better than LS estimation and "MMSE sim" estimation. Thus, SBGRU estimator shows the performance robustness with the different pilot densities.

\section{Conclusion}
In this paper, a DL-based channel estimator is designed under the time varying Rayleigh fading channel. The proposed DL-based channel estimator can achieve better performance than traditional algorithms and some NN estimators with different structures. Besides, the proposed NN channel estimator shows its ability to dynamically track the channel and its robustness with pilot density. 
\par In the traditional communication, there are much more complex traditional algorithms to complete channel estimation. However, there are some unique advantages compared with the traditional algorithms when deep learning algorithms are used.
\begin{itemize}
	\item Despite many estimation methods having been developed in traditional communication system, most of them always assume the channel to be invariant in coherence time. However, using deep learning algorithm, the prior knowledge about channel model and the channel invariant in coherence time assumption aren't needed during the training and testing, which shows the potential performance of DL-based algorithm under the time varying channel.
	\item The channel estimator designed in this paper can be easily optimized by combining traditional algorithms. For example, it's convenient to insert the high performance channel coding before the modulation to protect the performance against Gaussian noise. Thus, the MSE performance can be further improved.
\end{itemize}
\par In addition, there is still a lot work to do in applying deep learning or machine learning technology to the physical layer under time varying channel and here are some following aspects.
\begin{itemize}
	\item Except for the channel estimation, it is also feasible to construct a detector to undertake the equalization and demodulation together using deep learning algorithm. Thus, by connecting NN estimator and NN detector, a wireless communication system can be constructed. It is worth to explore whether such DL-based system can achieve better bit error rate(BER) performance than traditional system under the time varying channel and is still robust with different pilot densities.
\end{itemize}


%





\ifCLASSOPTIONcaptionsoff
  \newpage
\fi



\bibliographystyle{IEEEtran}
\bibliography{IEEEabrv,bib/paper}

\end{document}